\documentclass[10pt,twocolumn,twoside,letterpaper]{IEEEtran}

\usepackage{graphicx}
\usepackage{caption}
\usepackage{subcaption}
\usepackage{geometry}
\usepackage{hyperref}
	
\makeatletter
\def\ps@IEEEtitlepagestyle{
  \def\@oddfoot{\mycopyrightnotice}
  \def\@evenfoot{}
}
\def\mycopyrightnotice{
  {\footnotesize
  \begin{minipage}{\textwidth}
  \centering
  978-1-6654-8872-3/22/\$31.00 \copyright2022 IEEE
  \end{minipage}
  }
}

\usepackage{url}

\begin{document}
\title{Developing a Monolithic Silicon Sensor in \\
a 65\,nm CMOS Imaging Technology \\
for Future Lepton Collider Vertex Detectors
}
%
%
%
\author{\textbf{A.~Simancas}$^{1,2}$, J.~Braach$^{3,4}$, E.~Buschmann$^3$, A.~Chauhan$^1$, D.~Dannheim$^3$, M.~Del~Rio~Viera$^{1,2}$, K.~Dort$^{3,5}$, D.~Eckstein$^1$, F.~Feindt$^1$, I.M.~Gregor$^1$, K.~Hansen$^1$, L.~Huth$^1$, L.~Mendes$^{1,6}$, B.~Mulyanto$^1$, D.~Rastorguev$^{1,7}$, C.~Reckleben$^1$, S.~Ruiz Daza$^{1,2}$, P.~Schütze$^1$, W.~Snoeys$^3$, S.~Spannagel$^1$, M.~Stanitzki$^1$, A.~Velyka$^1$, G.~Vignola$^{1,2}$, H.~Wennlöf$^1$.
\thanks{$^1$Deutsches Elektronen-Synchrotron DESY, Germany (corresponding author email: adriana.simancas@desy.de)}
\thanks{$^2$University of Bonn, Germany}
\thanks{$^3$ Conseil Européen pour la Recherche Nucléaire CERN, Switzerland}
\thanks{$^4$ University of Hamburg, Germany}
\thanks{$^5$ University of Giessen, Germany}
\thanks{$^6$University of Campinas, Brazil}
\thanks{$^7$University of Wuppertal, Germany}
\thanks{Manuscript received December 9, 2022.}
\thanks{The Tangerine project is funded by the Helmholtz Innovation Pool, 2021 - 2023.}
\thanks{This work has been sponsored by the Wolfgang Gentner Programme of the German Federal Ministry of Education and Research (grant no. 13E18CHA).} 
\thanks{This project has received funding from the European Union’s Horizon 2020 Research and Innovation programme under GA no 101004761.}}

\maketitle

\pagenumbering{gobble}

\begin{abstract}
Monolithic CMOS sensors in a 65\,nm imaging technology are being investigated by the CERN EP Strategic R\&D Programme on Technologies for Future Experiments for an application in particle physics. The appeal of monolithic detectors lies in the fact that both sensor volume and readout electronics are integrated in the same silicon wafer, providing a reduction in production effort, costs and scattering material. The Tangerine Project WP1 at DESY participates in the Strategic R\&D Programme and is focused on the development of a monolithic active pixel sensor with a time and spatial resolution compatible with the requirements for a future lepton collider vertex detector. By fulfilling these requirements, the Tangerine detector is suitable as well to be used as telescope planes for the DESY-II Test Beam facility. The project comprises all aspects of sensor development, from the electronics engineering and the sensor design using simulations, to laboratory and test beam investigations of prototypes. Generic TCAD Device and Monte-Carlo simulations are used to establish an understanding of the technology and provide important insight into performance parameters of the sensor. Testing prototypes in laboratory and test beam facilities allows for the characterization of their response to different conditions. By combining results from all these studies it is possible to optimize the sensor layout. This contribution presents results from generic TCAD and Monte-Carlo simulations, and measurements performed with test chips of the first sensor submission.
\end{abstract}


\section{Introduction}
\IEEEPARstart{M}{onolithic} CMOS sensors have found their way through imaging technologies into High Energy Physics thanks to their good performance in particle detection. The main characteristic is the integration of an active sensor and readout in the same silicon wafer, which provides a reduction in production effort, costs and scattering material.

The Tangerine Project (\textbf{T}ow\textbf{a}rds \textbf{N}ext \textbf{Ge}ne\textbf{r}ation S\textbf{i}lico\textbf{n} D\textbf{e}tectors)~\cite{Wennlof2022} aims to develop the next generation of low-material and high-precision pixel detectors using novel technologies, targeting the requirements of a vertex detector in a future lepton collider experiment. Work Package 1 at DESY collaborates closely with the CERN EP Strategic R\&D Programme~\cite{EPreport} Work Package 1.2 as well as the ALICE ITS3 Upgrade project~\cite{alice} in designing, simulating and testing prototypes in a 65\,nm CMOS Imaging Technology.

In the context of particle physics applications, this contribution presents the development of a \textit{monolithic active pixel sensor} (MAPS) with a small collection electrode in the aforementioned technology. Compared to the previously investigated 180\,nm process, the 65\,nm technology offers a significant improvement in the logic density of the pixels and a reduction of the material thickness. The small collection electrode sensor is characterized by a low input capacitance (in the order of fF), granting a high signal to noise ratio (S/N) and a low power consumption.

The goal of the project is to develop a detector suitable for vertex-finding at future lepton colliders. Therefore, the performance targets are a position resolution of $\sim$3\,µm, a time resolution of the order of nanoseconds and a material budget of maximally 50\,µm of silicon. Each pixel will provide a charge measurement making it possible to perform charge-weighted position measurements. Due to similar performance requirements for telescope planes at test beam facilities, this detector will be suited to replace the current telescope planes at the DESY-II Test Beam facility~\cite{desy-tb}.

In the first foundry submission of the 65\,nm technology collaboration, test chips have been designed and produced. Joint contributions have taken place for their characterization, as well as generic simulations validating process modifications for enhanced sensor performance. A test chip featuring a Krummenacher type charge sensitive amplifier designed at DESY was characterized in~\cite{Feindt2022}. The ALICE Analog Pixel Test Structure (APTS) prototype~\cite{apts} is an analog prototype designed to characterize different sensor layouts and has been designed as part of the ALICE ITS3 upgrade efforts.

In this work, first sensor simulations as well as preliminary measurement results of the ALICE APTS are presented.

\section{SENSOR TECHNOLOGY}

Currently, three different sensor layouts are studied: \textit{standard}~\cite{Senyukov2013}, \textit{\textit{n}-blanket}~\cite{Snoeys2017} and \textit{\textit{n}-gap}~\cite{Munker2019}. These designs were originally developed in a 180\,nm CMOS imaging technology to enhance depletion, timing performance and radiation tolerance of small collection electrode MAPS.

\begin{figure}[t]
\centerline{\includegraphics[width=3.1in]{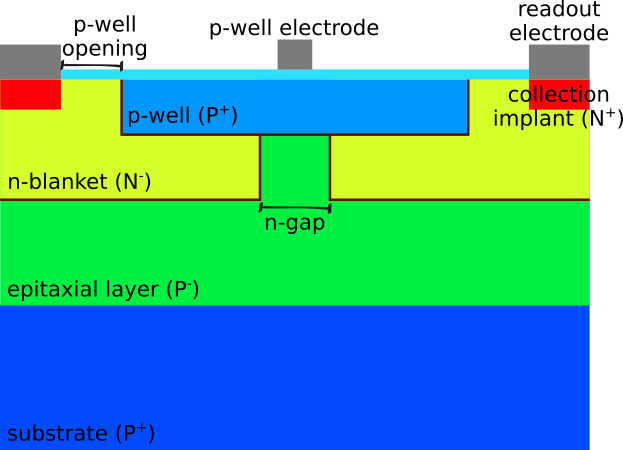}}
\caption{\textit{n}-gap layout example of the Tangerine sensor.}
\label{fig:scheme}
\end{figure}

Fig.~\ref{fig:scheme} is a schematic representation of a detector with the \textit{n}-gap layout. The cross section shows half of a pixel on each side of the schematic and the pixel edge in the center. A thin epitaxial \textit{p}-doped layer is grown on a low resistivity \textit{p}-doped substrate. The \textit{p}-well is the structure that hosts the in-pixel electronics and shields them from the electric field of the active sensor region. The \textit{n}-blanket is a low dose \textit{n}-doped layer implemented to create a planar \textit{pn}-junction, enlarging the depleted volume of the sensor. Furthermore, a gap in the \textit{n}-blanket produces a vertical \textit{pn}-junction that generates a lateral electric field in the farthest position from the readout electrodes (pixel boundaries).

To understand the 65\,nm CMOS Imaging Technology and optimize the sensor design, generic simulations and prototype testing are carried out simultaneously. The goal of the design optimization is to customize the electric field inside the sensor to improve efficiency and signal formation time. The simulation cycle is of the utmost importance to reduce time and costs invested in producing and testing prototypes.

\section{SENSOR SIMULATIONS}

Since the electric fields in small collection electrode MAPS are complex, device simulations are needed to provide insight into performance parameters of the sensor. Studies using generic doping profiles were performed for the three different layouts and for pitches between 10\,µm and 35\,µm. These simulations include only the epitaxial layer, since no significant contribution from the substrate to the electric field modelling is expected. Additionally, to bias the substrate in the TCAD simulations a metallic contact is added in the backside, as opposed to the the real detector that does not include a backside metalization.

\subsection{Technology Computer-Aided Design (TCAD) Simulations}
\label{sec:TCAD}

TCAD is a powerful tool to simulate electrical properties of semiconductors and can be used to optimize the sensor layout and other features, such as bias voltage configuration, to achieve the desired performance goals. TCAD contains a finite-element simulation tool that constructs a mesh over the studied structure and solves Poisson’s equation and carrier continuity equations to model the electrostatic potential and other properties in each node of the mesh. The software used for the simulations in this work is Sentaurus TCAD from Synopsys~\cite{Synopsys}.

\begin{figure}[t]
\centerline{\includegraphics[width=3.3in]{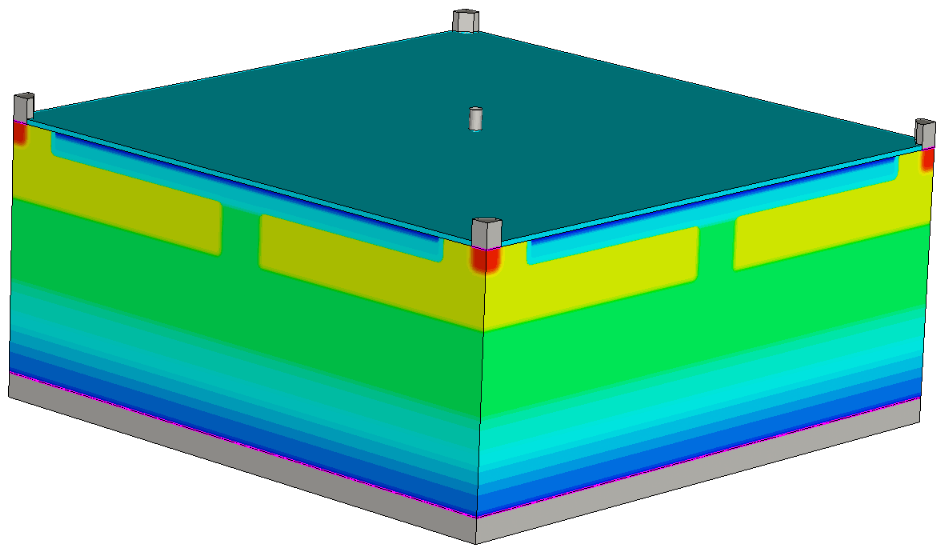}}
\caption{Generic 3D TCAD simulation of the \textit{n}-gap layout.}
\label{fig:3D}
\end{figure}

3D quasi-stationary simulations using generic doping profiles were performed to model the electric fields of the studied layouts, since 2D simulations cannot resolve some of the effects introduced by the lateral electric fields. An example of a generic 3D TCAD simulation for the \textit{n}-gap design is shown in Fig.~\ref{fig:3D}. The TCAD studies carried out so far include scans over different geometrical and operational parameters of the sensor, such as \textit{p}-well opening and bias voltage, and observing the behavior of the electric field, the lateral electric field strength, as well as the depleted volume. The results are reported in the following.

The \textit{\textit{p}-well opening} is the distance between the edge of the collection implant and the edge of the \textit{p}-well, as shown in Fig.~\ref{fig:scheme}. It was varied from 1\,µm to 4\,µm and from the results of the standard layout it was observed that increasing the \textit{p}-well opening provided a larger depleted volume and stronger lateral electric field. This translates into a larger signal and a faster signal formation, but reduces the space for readout electronics. However, this effect was much less prominent once the \textit{n}-blanket modification was added.

The \textit{gap size} in the \textit{n}-gap layout was varied from 1\,µm to 4\,µm. The gap creates a lateral electric field at the pixel boundaries, which increases with the gap size. The gap is introduced to allow for the free charges to avoid the field minimum reported in \cite{Munker2019} and drift with a shorter mean free path towards the collection electrodes, thus it improves the signal formation time.

\begin{figure}[tbp]
\centering
\begin{subfigure}[b]{.5\textwidth}
  \centering
    \includegraphics[width=3in]{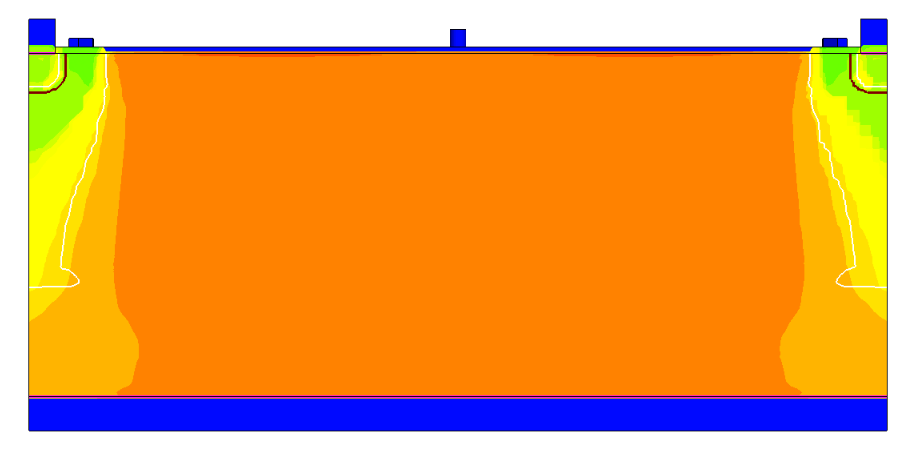}
  \caption{Standard}
  \label{fig:bias_std}
\end{subfigure}
\begin{subfigure}{.5\textwidth}
  \centering
    \includegraphics[width=3in]{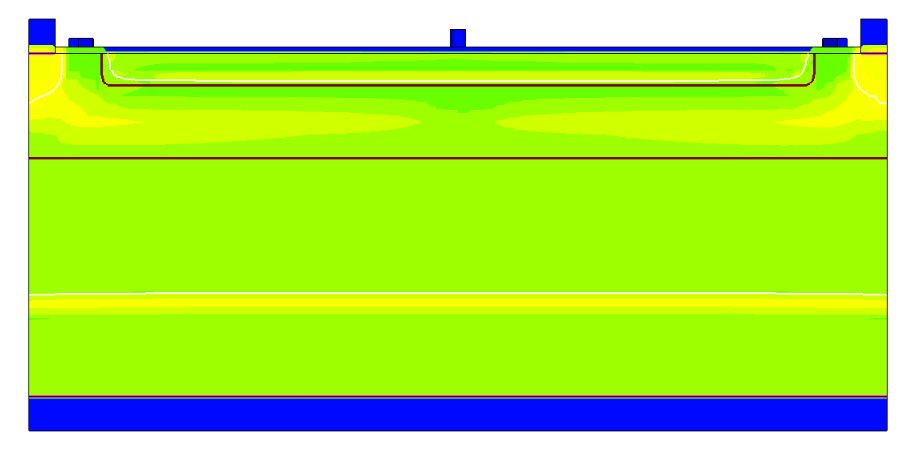}
  \caption{\textit{n}-blanket}
  \label{fig:bias_blk}
\end{subfigure}
\begin{subfigure}{.5\textwidth}
  \centering
    \includegraphics[width=3in]{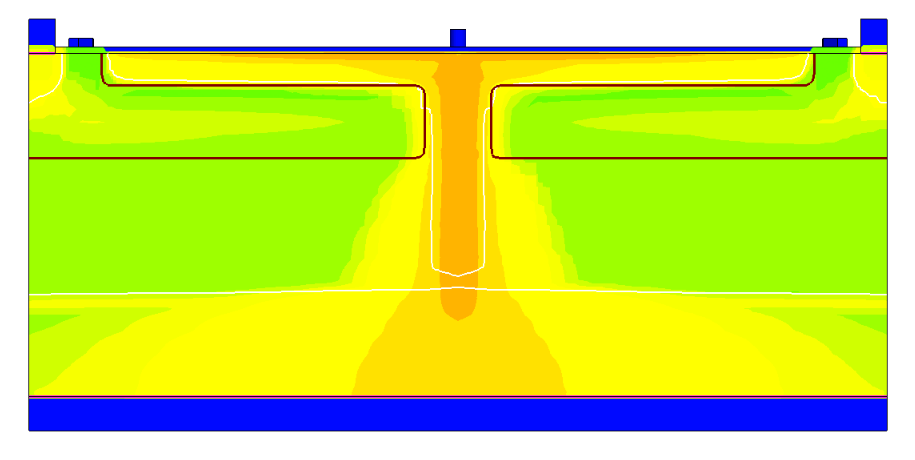}
  \caption{\textit{n}-gap}
  \label{fig:bias_gap}
\end{subfigure}
\caption{Generic TCAD Simulation: current density for the three layouts with \textit{p}-well bias of -1.2\,V and substrate bias of -4.8\,V. Pixel pitch of 25\,µm. Brown line indicates the position of the \textit{pn}-junction and white line corresponds to the depleted volume. The high current density in the edge of the pixels for the standard and the \textit{n}-gap design indicates breakdown for these biasing conditions.}
\label{fig:bias}
\end{figure}

A \textit{breakdown} of the sensor was observed when fixing the \textit{p}-well bias at -1.2\,V, and applying a higher bias to the substrate. The breakdown was detected early on for the standard layout, at -2.4\,V, for the \textit{n}-blanket layout it was reached at a bias of -11\,V, and for the \textit{n}-gap layout it was observed at -4.8V. This behavior is in agreement with observations made during monitoring of the leakage current in experimental measurements. Fig.~\ref{fig:bias} shows how the breakdown is modelled in the generic TCAD simulation. The color scale corresponds to the current density, while the white line indicates the limits of the depleted volume. The breakdown produces a high current density in the edge of the pixels, and the depleted volume is deformed.

The \textit{bias voltage} applied to the \textit{p}-well and the substrate was scanned from 0\,V to -20\,V, while fixing the bias of the readout electrodes to 1.2\,V. When the bias voltage was simultaneously increased for both electrodes, an increased depleted volume was observed. However, high values of the electric field were detected inside the \textit{p}-well structure, compromising the shielding of the electronics. When the bias was increased only for the substrate while fixing the \textit{p}-well bias, a similar behaviour was observed for the \textit{n}-blanket and \textit{n}-gap layouts, but the \textit{p}-well integrity was preserved.

\begin{figure}[tbp]
\centering
\begin{subfigure}[b]{.5\textwidth}
  \centering
    \includegraphics[width=3in]{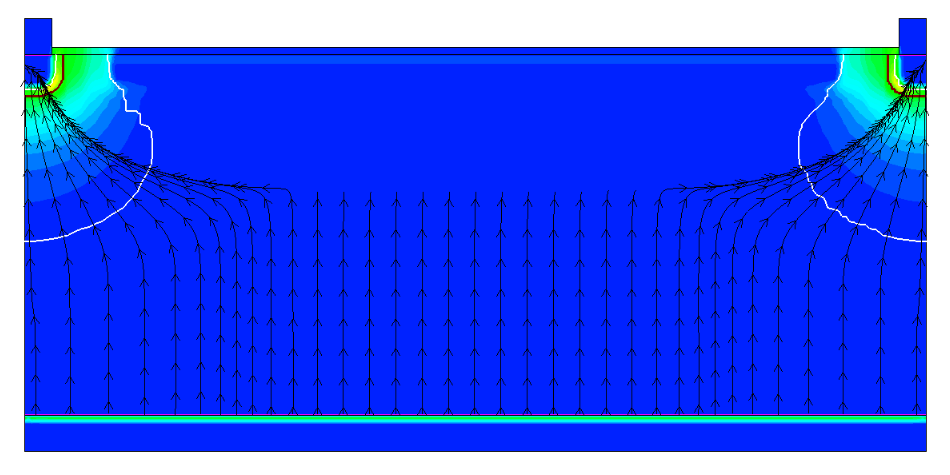}
  \caption{Standard}
  \label{fig:EF_std}
\end{subfigure}
\begin{subfigure}{.5\textwidth}
  \centering
    \includegraphics[width=3in]{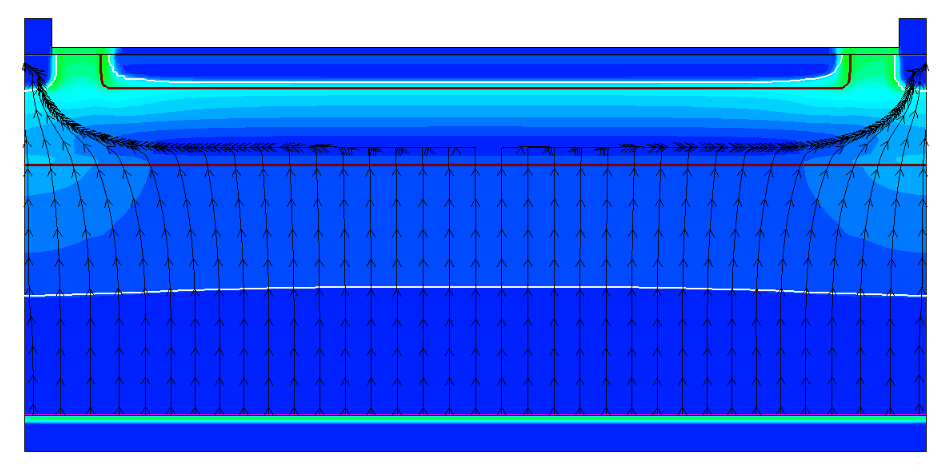}
  \caption{\textit{n}-blanket}
  \label{fig:EF_blk}
\end{subfigure}
\begin{subfigure}{.5\textwidth}
  \centering
    \includegraphics[width=3in]{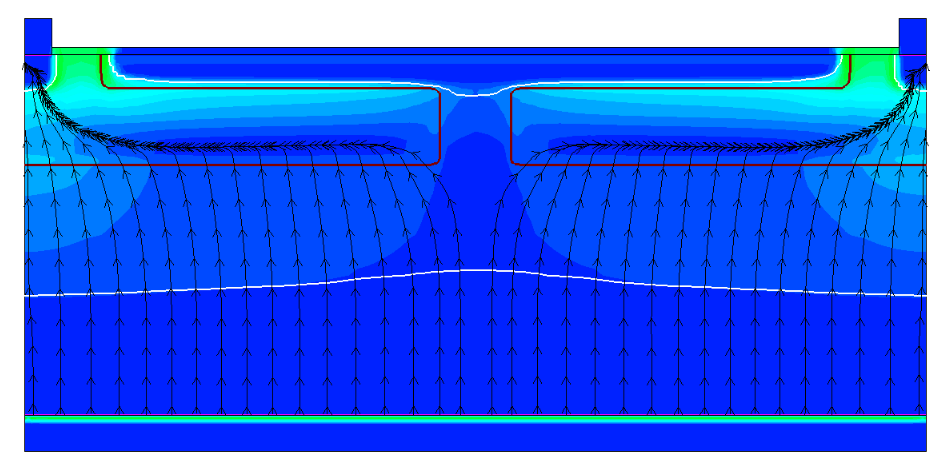}
  \caption{\textit{n}-gap}
  \label{fig:EF_gap}
\end{subfigure}
\caption{Generic TCAD Simulation: electric field for the three layouts with \textit{p}-well and substrate bias of -4.8\,V. Pixel pitch of 25\,µm. Brown line indicates the position of the \textit{pn}-junction, black arrows represent the streamlines and the white line delimits the depleted volume.}
\label{fig:EF}
\end{figure}

Fig.~\ref{fig:EF} shows the electric field obtained from TCAD simulations of the three sensor layouts with generic doping profiles, where \textit{p}-well and substrate bias voltage were set to -4.8\,V. The brown line indicates the position of the \textit{pn}-junction, the white line delimits the depleted region and the streamlines (black arrows) indicate the instantaneous tangent to the velocity vector of the moving charges.

Comparing between the different layouts with a pixel pitch of 25\,µm, the following can be concluded:
\begin{itemize}
    \item The standard layout (Fig.~\ref{fig:EF_std}) has a small depleted volume. The electron-hole pairs produced outside the depleted volume will move predominantly by diffusion in random directions and some might not reach the readout electrodes. The expected effect on charges produced at the edge of the pixel is a low efficiency, but a high charge-sharing between pixels, which improves the spatial resolution.
    \item The \textit{n}-blanket layout (Fig.~\ref{fig:EF_blk}) shows a larger depleted volume. The electron-hole pairs produced in the active volume will move predominantly by drift towards the readout electrodes. This is foreseen to produce an improvement in efficiency, but with an impairment in spatial resolution due to lower charge-sharing.
    \item The \textit{n}-gap layout (Fig.~\ref{fig:EF_gap}) has a higher lateral electric field in the pixel edges. The electron-hole pairs produced in the edge of the pixel will
    drift with a shorter mean free path towards the readout electrodes. As a consequence, an improvement is expected in efficiency as well as the signal formation time, but with a further impairment in spatial resolution due to even lower charge-sharing.
\end{itemize}

In order to quantify the effects discussed here, Monte-Carlo simulations are required as in Section \ref{sec:MC}.

TCAD is also capable of simulating current pulses produced by the interaction of a charged particle with the sensor. This is carried out with \textit{transient simulations} and can be used to estimate signal characteristics, such as signal magnitude and time evolution.

\begin{figure}[tbp]
\centerline{\includegraphics[width=3.6in]{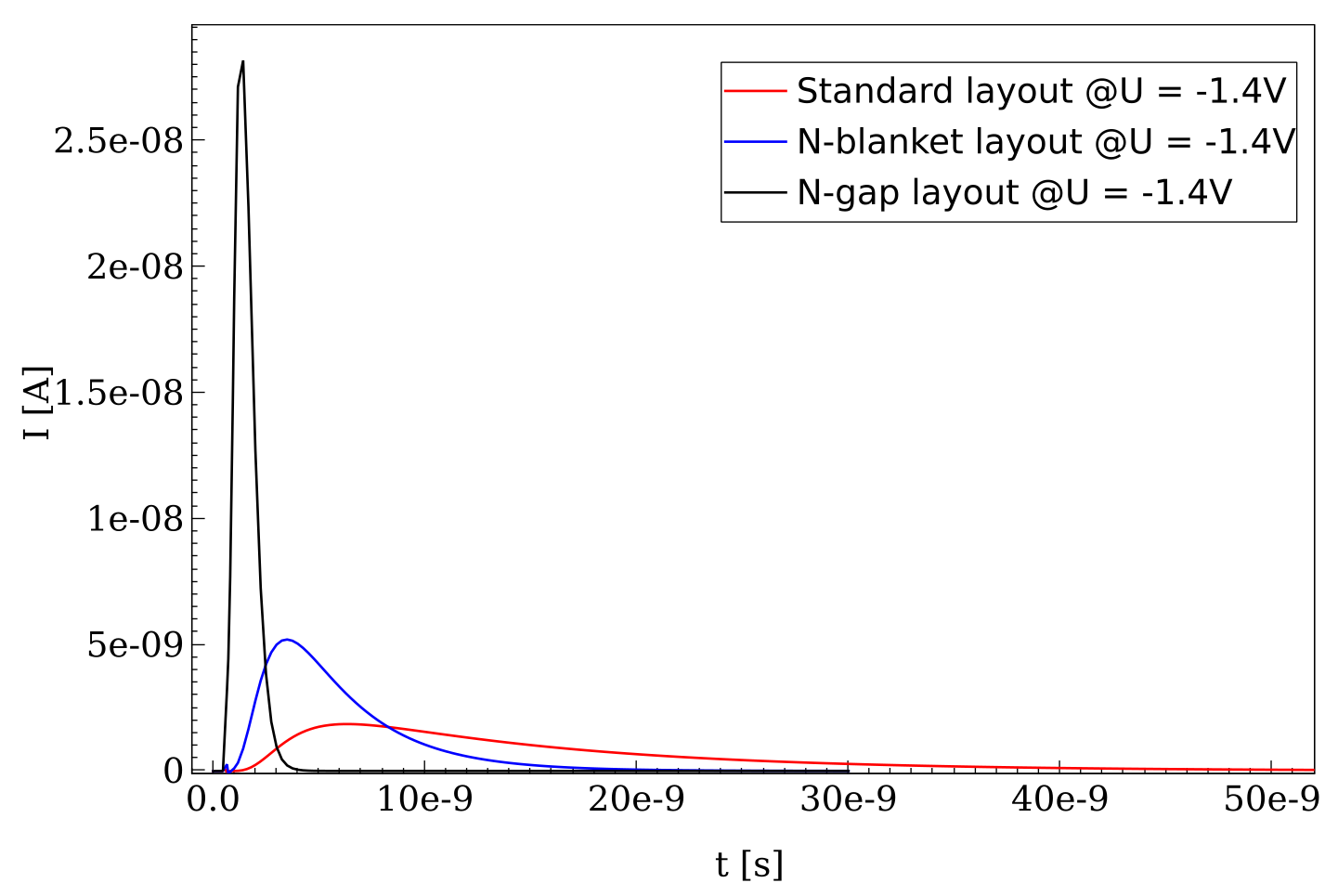}}
\caption{Generic TCAD Simulation: current pulse produced by a MIP traversing the corner of a pixel, for the three different layouts.}
\label{fig:trans}
\end{figure}

A transient simulation for a \textit{minimum ionizing particle} (MIP) traversing the corner of a pixel with an electron/hole pair production of 63\,eh/µm was performed for the studied layouts. The result is shown in Fig.~\ref{fig:trans}, confirming that the time evolution of the signal is improved by the modifications in the sensor, and particularly for the \textit{n}-gap layout.

For a standard layout sensor, a similar simulation was performed with a MIP traversing the center of the pixel. By integrating the induced charge for the duration of the signal, a total of $\sim$750 electrons was obtained. Furthermore, generic TCAD transient simulations have provided valuable feedback for the ASIC design, such as expected signal magnitudes to define a reasonable threshold in the readout electronics.

\subsection{Allpix$^2$ Simulations}
\label{sec:MC}

TCAD transient simulations are time consuming and have a high computational budget, thus quasi-stationary simulations from TCAD are combined with Monte-Carlo simulations to obtain high statistics data and calculate the performance parameters of the sensor~\cite{tcad+mc}. Within the Tangerine project, this combination of simulations is used to quantify the effects reported in Section \ref{sec:TCAD}.

Allpix$^2$~\cite{allpix} is a modular framework developed for Monte-Carlo simulations of semiconductor radiation detectors. It provides the possibility to build a matrix of pixels by replicating the TCAD electric fields simulated on a single cell.

First results from Monte-Carlo simulations using generic TCAD fields from this work have been reported in~\cite{Wennlof2022}. The performance parameters of interest are detection efficiency, cluster size, and spatial resolution. Results confirm that the modifications to the sensor layout are valuable as they increase the efficient operating margin of the sensor. The observed trends are equivalent for all tested bias voltages.

\section{PROTOTYPE TESTING}

The purpose of the ALICE APTS~\cite{apts} prototype is to characterize different sensor designs. Several of these chips are being studied and the results will allow for direct comparisons with simulations to be performed. This section presents the activities within the Tangerine project of system integration and testing of the ALICE APTS in laboratories and test beam campaigns with preliminary results.

\subsection{APTS and Data-Acquisition (DAQ) System}
\label{sec:APTS}

 The ALICE APTS contains a matrix of 4 × 4 square pixels and has been produced in the three studied sensor designs. The tested devices have a pixel pitch of 25\,µm, the pixels are DC coupled to the front-end electronics and each pixel contains a source follower as buffered analog output. The latter makes the APTS a structure suitable for comparisons against simulations.

 The Caribou System~\cite{caribou1, caribou2} is used as data-acquisition system for the ALICE APTS. It is an open-source set of hardware, firmware and software for laboratory and beam tests. The modular hardware consists of three boards: the application-specific \textit{chip board} containing the sensor, the periphery \textit{CaR board} which provides current and voltage sources together with a physical interface between System-on-Chip (SoC) and the detector, and the \textit{evaluation board} that contains the Xilinx Zynq SoC, which runs the detector control and the data-processing firmware.

For the ALICE APTS, a custom chip board was designed and produced with amplification and signal shaping at each pixel output. The readout was performed with two 8-channel 65\,MS/s ADCs on the CaR board, using a custom firmware block. The firmware provides a configurable trigger logic, where either an external or an internal trigger can be selected with programmable thresholds and adjustable latency. The firmware is compatible with the AIDA Trigger Logic Unit~\cite{tlu} (TLU).

The data-acquisition framework employed in the project is EUDAQ2~\cite{eudaq2}. It is a generic data-acquisition software for use in conjunction with beam telescopes at charged particle beam lines. It allows for storage and synchronization of data from several systems. A detector-specific decoder is used to interpret the raw data for further analysis.

\begin{figure}[tbp]
\centerline{\includegraphics[width=3.5in]{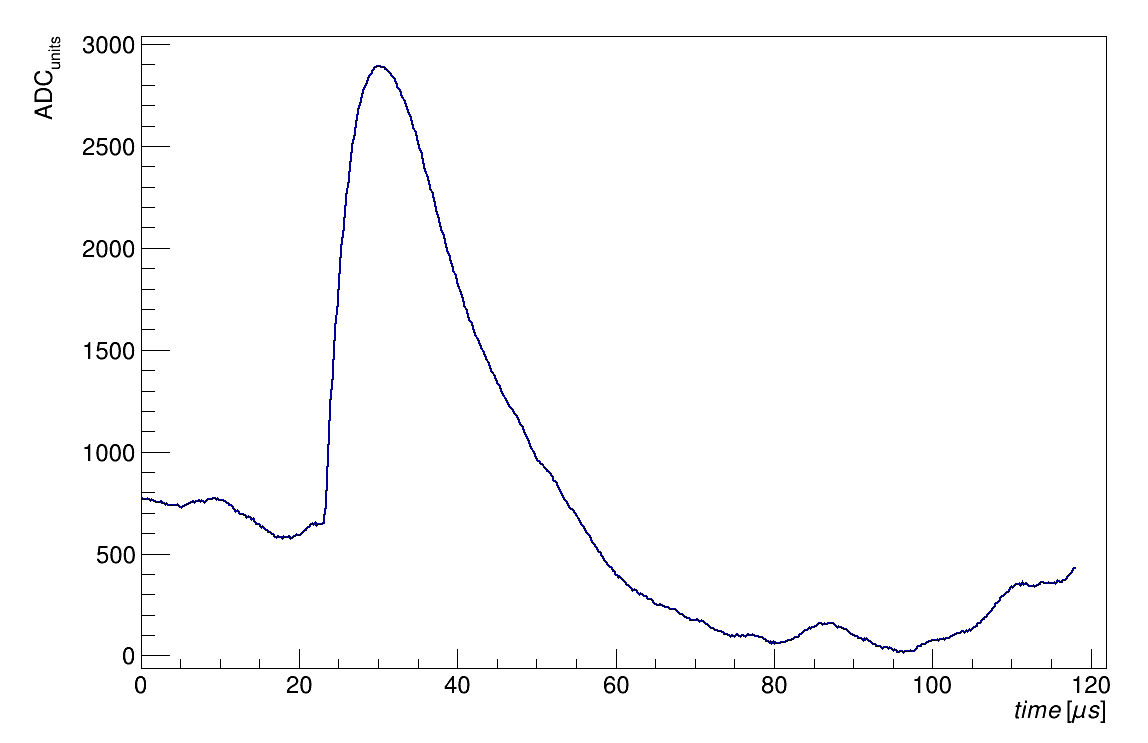}}
\caption{Waveform of a MIP measured with the ALICE APTS in the \textit{n}-blanket layout at -3.6\,V \textit{p}-well and substrate bias.}
\label{fig:wf}
\end{figure}

The recorded data with this DAQ system consists of waveforms sampled at 65 MHz with a configurable number of samples. Internal triggers were used for laboratory tests, while for the test beam data acquisition an external trigger was provided by the TLU. Fig.~\ref{fig:wf} shows a waveform example produced by a MIP in an APTS with the \textit{n}-blanket layout. The shape of the waveform is dominated by the amplifiers employed in the chip board of the DAQ system.

\subsection{Laboratory Characterization}
\label{sec:lab}

The activities performed in the laboratory involved the optimization of the front-end operation parameters and studies with charge injection and radioactive sources for gain calibration.

Since the acquired signals are represented in ADC units, a calibration is required to relate these values to number of electrons. This way, it is possible to match the thresholds of data and simulation and quantify the agreement between them. The calibration of the tested devices was performed with X-ray fluorescence, an $^{55}$Fe radioactive source and test pulses.

\begin{figure}[tbp]
\centerline{\includegraphics[width=3.6in]{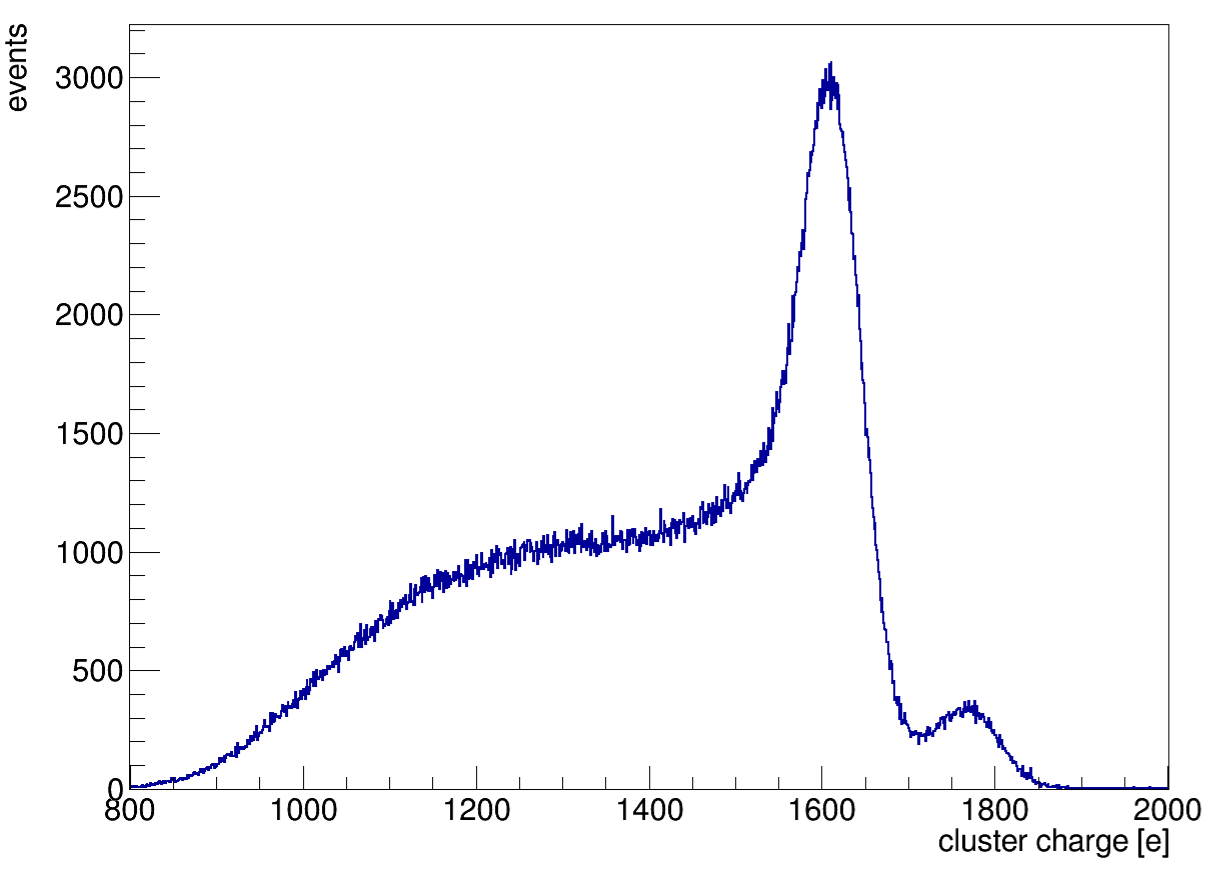}}
\caption{Calibrated spectrum of an $^{55}$Fe source acquired with an ALICE APTS in the \textit{n}-blanket layout at -3.6\,V \textit{p}-well and substrate bias.}
\label{fig:Fe55}
\end{figure}

The decay of $^{55}$Fe produces two X-ray emissions that are considered monochromatic: K-alpha of 5.9\,keV and K-beta of 6.5\,keV. For the interaction of these X-rays with silicon, a production of approximately 1640 and 1800 electron-hole pairs, respectively, is expected. The calibration uses the K-alpha and K-beta peaks shown in the $^{55}$Fe spectrum in Fig.~\ref{fig:Fe55}, which was measured with an APTS in the \textit{n}-blanket layout. The result is a combination of the individual pixel charge spectra, so the tail at the left of the spectrum is a combination of Compton scattering, charge-sharing and threshold effects that differ per pixel. The calibration was validated by applying the calibration on the spectrum from the X-ray fluorescence of titanium.

\subsection{test beam Measurements}
\label{sec:TB}

The ALICE APTS was tested in the DESY-II Test Beam Facility~\cite{desy-tb} to characterize its performance with MIPs. A MIMOSA26 telescope~\cite{mimosa} was used as a reference system to reconstruct the individual particle tracks of the beam. The APTS acted as a \textit{Device Under Test} (DUT) and was placed in the middle of the 6 telescope planes, orthogonal to the beam. Additionally, a TelePix~\cite{telepix} plane was used as trigger in the data acquisition. It provides time stamps with a precision below 5\,ns, and a fast digital hit output signal which allows for triggering in a configurable region of interest. This plane was placed downstream with respect to the telescope. With this setup it is possible to determine the effect of particles impinging on different positions in the evaluated prototype and obtain the performance parameters mentioned in Section \ref{sec:MC}.

For the data analysis, a modular framework called Corryvreckan~\cite{corry} is used. It allows for online monitoring and offline event building in complex data-taking environments combining detectors with different readout architectures. With this software, the particle tracks are reconstructed from the information provided by the telescope planes. One particle interaction can produce a signal in different pixels. The one that registers the higher signal is called \textit{seed pixel}, and together with the surrounding pixels they constitute a \textit{cluster}. After correlating the tracks with the DUT information, clusters in the DUT are associated to the reconstructed tracks and it is possible to perform efficiency and resolution studies.

The devices were tested under different operational conditions. Given the maximum beam rate of 5\,kHz, a beam cross section of the order of centimeters and the active area of the 25\,µm pitch ALICE APTS of 100 × 100\,µm, it was necessary to record data for several hours for each investigated setting. It was discovered that, due to the long acquisition time and temperature changes in the test beam area, the relative position of the DUT with respect to the reference was changing. This required to develop a new alignment method that corrected the relative drift within the same acquisition run. This method has shown promising results, but it is still under evaluation. For this reason, the quantitative results are still preliminary, but it has been possible to observe encouraging qualitative results that are reported in the following.

\begin{figure}[tbp]
\centerline{\includegraphics[width=3.7in]{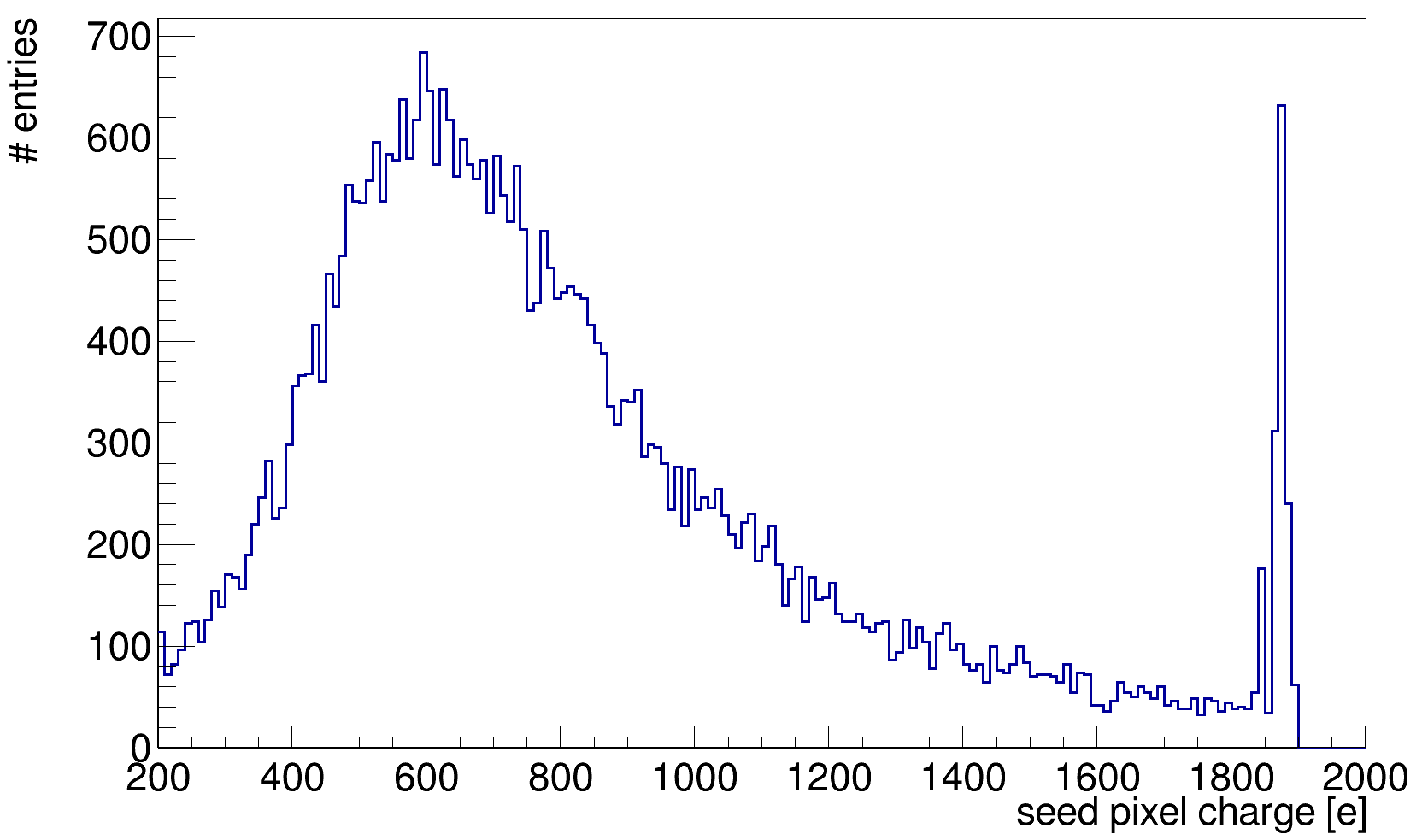}}
\caption{Charge distribution of seed pixels for associated clusters acquired with an ALICE APTS in the \textit{n}-blanket layout at -3.6\,V \textit{p}-well and substrate bias.}
\label{fig:charge}
\end{figure}

After applying the calibration to test beam data of an \textit{n}-blanket design, the charge distribution of the seed pixels was obtained as shown in Fig.~\ref{fig:charge}. The distribution corresponds approximately to a Landau distribution convolved with a Gaussian. A \textit{most probable value} of around 600 electrons is observed. The peak at the end of the distribution is due to saturation of the ADCs.

The \textit{cluster size} depends on the charge-sharing between pixel cells and the chosen thresholds. The detectors with sensor modifications (\textit{n}-blanket and \textit{n}-gap layouts) exhibited a lower cluster size in comparison to the standard layout, indicating a similar trend to the observations made in simulations.

The \textit{detection efficiency} of the detector was measured by relating the particle hits in the DUT to the reconstructed tracks. As expected from simulations and previous works~\cite{Snoeys2017,Dyndal2020}, an improvement in efficiency was observed for the \textit{n}-blanket and \textit{n}-gap layouts.

\section{Conclusion}

The Tangerine project participates in the investigation of monolithic pixel detectors in 65\,nm CMOS Imaging Technology lead by the CERN EP Strategic R\&D Programme on Technologies for Future Experiments. 

A monolithic active pixel sensor in 65\,nm CMOS imaging technology is being developed within the Tangerine project. Simulations and prototype testing are complementary activities carried out within the project and the collaboration institutes.

Device simulations using generic doping profiles provide valuable insight for sensor optimization and understanding sensor behaviour for novel technologies. The combination with Monte-Carlo simulations produces results which can be directly compared with experimental tests.

Generic TCAD simulations have confirmed that the principles to modify the sensor in the 180\,nm and 65\,nm are generally applicable. Generic simulations have allowed for establishing the parameters of relevance for the sensor optimization and their effect on the operation of the detector, such as \textit{p}-well opening, \textit{n}-gap size and sensor bias voltage. The results show an overall agreement with what has been observed in other technologies investigated for MAPS. Efficiency and resolution studies using generic TCAD simulations combined with Monte-Carlo simulations are ongoing.

The ALICE APTS has been evaluated in laboratory and test beams. A custom DAQ system based on Caribou was developed and integrated for these measurements. From the experimental activities, the functionality of the test setup has been demonstrated. Preliminary results on charge distribution, cluster size and detection efficiency have shown a qualitative agreement with simulations. More detailed studies including spatial
resolution and timing are continuing.

Further test campaigns are planned for the near future, as well as dedicated simulations on timing performance. A fully integrated chip with a larger pixel matrix, designed jointly by CERN, DESY and IFAE, has been submitted to the foundry. This will allow for recording high-statistic data samples and for
further improvement of the comparison with simulations.

\section*{Acknowledgments}

The measurements presented have been performed at the test beam Facility at DESY Hamburg (Germany), a member of the Helmholtz Association (HGF).

The authors wish to express their gratitude to the CERN EP R\&D WP 1.2 and especially to the designers of the APTS and the ALICE ITS3 measurement team for their support.

(c) All figures and pictures by the author(s) under a \href{https://creativecommons.org/licenses/by/4.0/}{CC BY 4.0} license, unless otherwise stated.



%

\bibliography{ieee22_biblio_new} 
\bibliographystyle{IEEEtran}




\end{document}